\documentclass[twocolumn,aps,prl,groupedaddress]{revtex4}
\usepackage{graphicx}
\usepackage{dcolumn}
\usepackage{bm}

\begin{document}
\bibliographystyle{apsrev}
\title{Quantum statistical effects in nano-oscillator arrays}

\author{Douglas M. Photiadis}
\email{douglas.photiadis@nrl.navy.mil}
\author{J.A. Bucaro}
\author{Xiao Liu}
\affiliation{Naval Research Laboratory, 4555 Overlook Ave. SW, Washington, DC 20375-5320}
%

\pacs{}

\begin{abstract}
{
We have theoretically predicted the density of states(DOS), the low temperature specific heat, and Brillouin scattering spectra of a large, free standing array of coupled nano-oscillators.  We have found significant gaps in the DOS of 2D elastic systems, and predict the average DOS to be nearly independent of frequency over a broad band $f < 50GHz$.  At low temperatures, the measurements probe the quantum statistics obeyed by rigid body modes of the array and, thus, could be used to verify the quantization of the associated energy levels.  These states, in turn, involve center-of mass motion of large numbers of atoms, $N \gtrsim 10^{14}$, and therefore such observations would extend the domain in which quantum mechanics has been experimentally tested.  We have found the required measurement capability to carry out this investigation to be within reach of current technology.
}
\end{abstract}

\maketitle

	Over the past decade, significant effort has been focused on the observation of quantum effects in large systems, both to test quantum mechanics in this regime and to directly observe quantum decoherence phenomena arising from coupling to the environment.  Mechanical oscillators have often been considered in this context because displacement states of the oscillator can serve as ``Schrodinger cat'' states in the investigation of these phenomena.  The measurement concepts reported to date in this area\cite{Rego_quantized, Schwab_quantized, Armour-prbwhichpath, Carr_Crossover, Tittonen_interf}  have primarily involved investigating the quantum properties of a single oscillator, though short linear chains of coupled oscillators have been considered recently in the context of minimum thermal conductance\cite{Cleland-conductance} and entanglement\cite{Eisert-linearrays}.  Concepts involving a mechanical oscillator coupled to the optical modes of a cavity have been considered by Bose et. al.\cite{bose-pra99}, while the investigation of a mechanical oscillator coupled to the electrons in a Cooper pair box\cite{Armour-prl2002}, the electronic analogy, is underway.    Recent experimental results have pushed close to the limits of the uncertainty principle\cite{Schwab_and_Lahaye}.

A different approach, and the central focus of this paper, is to explore the extent to which one can probe the quantum behavior of mechanical oscillators by observing the effectively bulk properties of a large, free standing array of coupled nano-oscillators.   
Measurements of this type avoid the significant experimental difficulties that direct observations of the motion of individual nano-mechanical oscillators involve\cite{Schwab_quantized, Schwab_and_Lahaye, bose-pra99} and offer a new avenue to explore their quantum behavior.
The simplest bulk measurements do not explore the coherent quantum behavior of the system, but they do enable us to verify that macroscopic whole body modes obey quantum statistics and therefore have quantized energy levels.

The quantum behavior of the system is expected to be governed by the field operators
\begin{eqnarray}
\label{field_op}
\psi (x,t) =& \int [d\alpha ]& \left [ {\hbar \over 2 \rho \omega_\alpha} \right ]^{1/2} 
\left [ a_\alpha \exp (-i \omega_\alpha t ) \psi_\alpha (x) \right . \nonumber \\
&+& \left . a^\dag_\alpha \exp (+i \omega_\alpha t ) \psi^\ast_\alpha (x) \right ]
\end{eqnarray}
where the $a^\dag_\alpha (a_\alpha )$ are the usual creation (annihilation) operators for mode $\alpha$, $\rho$ is the mass density, and the $\psi_\alpha (x)$ are the classical eigenfunctions.  We have verified this intuitive result via the canonical quantization prescription, but the details are too lengthy to be given here; for our purposes in this letter, we assume  Eq. (\ref{field_op}) to be valid.  

The classical solution for the eigenfunctions and the spectrum is evidently central to the quantum behavior of the system (e.g. the presence and size of frequency gaps and the nature of the eigenfunctions).   Typically, the dynamic behavior of even a simple elastic structure contains significant complexity beyond a model of coupled simple oscillators\cite{liu-apl01}.  Therefore, below, we give numerical results for the classical modes of a generic, periodic, nano-structured system.  We have found large gaps in the spectrum over a broad frequency range, .1GHz $< f <$ 50GHz.  We then obtain theoretical predictions for two of the most basic bulk  measurements of such a system, the specific heat and Brillouin scattering.  At fairly low temperatures, $T \sim 30 mK$, the quantum statistical behavior of modes involving whole body motion of the nano-oscillator array is observable.

The geometry we consider is that of a periodic, planar structure as shown in Fig. 1.  The particular dimensions chosen can, of course, be rescaled by a factor $\lambda$ according to $L \rightarrow \lambda L$, $\omega \rightarrow \lambda^{-1} \omega$.
\begin{figure}[h]
\begin{center}\leavevmode
\includegraphics[width=.6\linewidth]{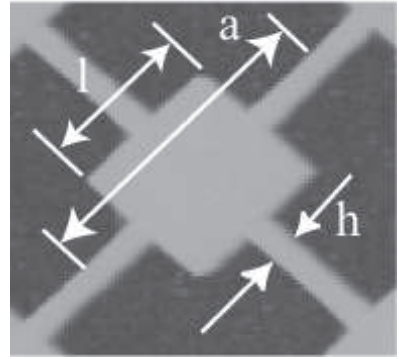}
\end{center}
\caption{
Geometry of a periodic planar nano-structured phononic crystal.  The values used for calculations are: $l = 200$nm, $a=360$nm, $h = 20$nm, thickness = 20nm, and material parameters of Si.}\label{fig1}\end{figure}
We assume, for simplicity, that the material is isotropic. 
The normal modes are calculated by seeking solutions of the Bloch form $\psi (x) = \exp(ikx) U(x)$,
where $k = (k_x, k_y) = (2n\pi / L_x , 2m\pi / L_y)$, and $U(x)$ is assumed to possess the periodicity $a$ of the lattice.  Here, $n$ and $m$ are integers, $L_x$ and $L_y$ are the dimensions of the crystal, and the boundary conditions are taken to be periodic, an approximation which will not be significant for a large system.  This results in a boundary value problem in the unit cell 
which yields the resonance frequencies for a particular wavenumber, with one frequency for each sub-band of the spectrum.  

 \begin{figure}[b]
\begin{center}\leavevmode
\includegraphics[width=.8\linewidth]{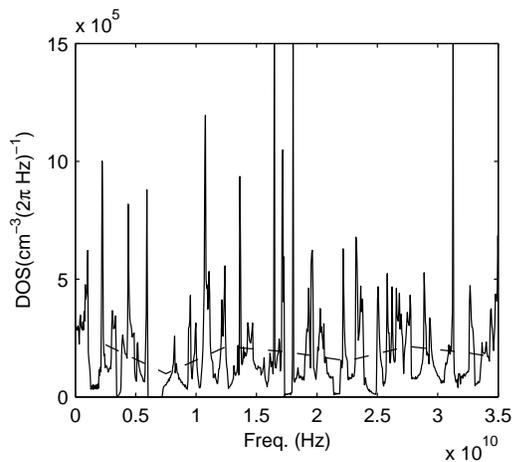}
\end{center}
\caption{
Density of phonon states of a periodic planar nano-structured array.  Solid: Total DOS; dashed: average DOS.}\label{fig2}\end{figure}
The numerically computed density of states(DOS) is shown in Fig. 2. The vibrational modes that execute whole body motion locally are the modes of greatest interest with regard to testing the limits of quantum mechanics\cite{leggett_qmlimits} and environmentally induced decoherence\cite{caldeira+Leggett-pra85, Walls-1985}, because the associated energy eigenstates exhibit coherent center of mass motion of large numbers of atoms.  The states in the lowest frequency band extending up to about 2GHz are in large part whole body modes involving significant domains of the entire structure.  The first quasi-gap in the spectrum, associated with weakly coupled local 1-1 modes (the peak at $\approx$ 2.25GHz), occurs in the range from 1-2GHz, while a true gap appears in the spectrum at approximately 6GHz.  

The presence of gaps is significant because modes near the band edges are Anderson localized by irregularity\cite{sheng-john_localization}.  This phenomena may be advantageous in the experimental study of the quantum behavior of the system because of the significantly larger physical displacements for the same energy levels.  Further, localized wave functions vanish exponentially on the boundaries and therefore promise to suffer much less decoherence resulting from coupling to the environment.

A notable feature of the results is that the average DOS is more or less constant in this frequency range.  This behavior  results from the dispersive properties of the flexural modes which dominate the spectrum because of their slow phase velocity.  One may see this qualitatively by estimating the DOS of a uniform, flexural wave system.  

The dispersion relation for flexural waves propagating in a thin plate of thickness $t$ is $\omega = (D/m)^{1/2} k_f^2$, with $D = Et^3/[12(1-\nu^2)]$ the flexural rigidity, and $m = \rho t$ the areal density.  Here $E$ is the Young's modulus, $\rho$ the density, $c_f = \omega /k_f$ the flexural wave speed, and $\nu$ the Poisson ratio.  The DOS of a uniform, d-dimensional, flexural wave system of size $L^d$ is therefore given by
\begin{equation}
\label{plate-DOS}
N(\omega ) 
= { \left ( \omega L/ 2 c_f \right )^{d} \over  \pi^{d/2} \Gamma (d/2) \omega } \stackrel{d \to 2}{=} 
{ L^2 \over 4\pi} (m/D)^{1/2}
\end{equation}
and is independent of frequency in two dimensions.  

Theoretical predictions for the specific heat can be obtained directly from the DOS/vol, $n(\omega )$, via
\begin{equation}
\label{spec_heat_fund}
c_v = {\partial u \over \partial T} = {k_b \over V} \int d\omega \ n( \omega ) {e^{x(\omega )} \over (e^{x(\omega )} - 1)^2}
\end{equation}
where $x(\omega ) = \hbar \omega /k_b T$.
The result is shown in the temperature range below 250mK in Fig. 3a.  In this range, the specific heat depends on the temperature in an approximately linear fashion.  This behavior differs from the typically quoted $T^2$ dependence\cite{Roukes_Yocto} of the specific heat in 2-D phonon systems because of the roughly constant DOS at low frequencies, an aspect discussed above.   

\begin{figure}
 \begin{center}\leavevmode
\includegraphics[width=.8\linewidth]{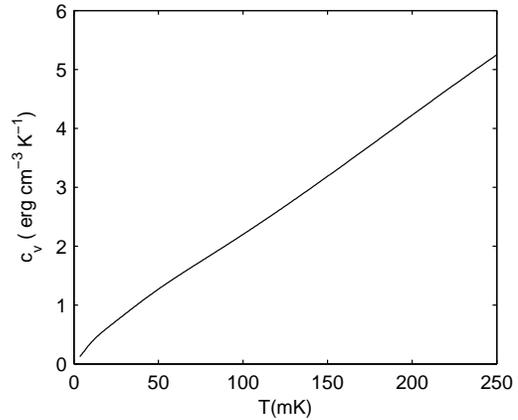}
\end{center}
\caption{
Specific heat of a periodic planar nano-structured array.}
\label{fig3}\end{figure}
Employing Eq. (\ref {plate-DOS}), one may show that the low temperature heat capacity of a uniform d-dimensional flexural system of size $L^d$ is
\begin{equation}
\label{plate_sh}
C_v  = k_b {L^d \over 2^d} \left ( {m^{1/2} k_b T \over D^{1/2} \pi \hbar} \right )^{d/2} 
{d\over 2}\left({d\over 2} + 1\right)\zeta (d/2 +1) ,
\end{equation}
where $\zeta$ is the Riemann zeta function.  In 1-D, the heat capacity varies as $T^{1/2}$, while in 2-D the predicted temperature dependence is $T$, as we have found above.  The heat capacity of low dimensional phonon systems is therefore larger at low temperatures than predicted using the customary $T^2$ temperature dependence of $C_v$.

The nano-structure of the system, as opposed to the overall dimensionality, has a relatively weak impact on the temperature dependence of the specific heat, giving rise to the bump in the vicinity of $T\sim 35mK$.  Nevertheless, the observation of a linear temperature dependence at low temperatures would provide direct evidence that the relevant whole body modes are behaving as simple quantum oscillators obeying Bose statistics.  

Experimentally, a  $1000 \times 1000$ array of oscillators with dimensions as given in Fig.1 would result in a heat capacity of $1.6\times 10^{-16}$J/K
at 100 mK, an estimate which does not include the addenda (heater and thermometer)
necessary for the measurement. This is a challenge to the current limit of
$2\times 10^{-9}$J/K at 1.5K \cite{Denlinger, Fominaya}.  But with the rapid advance of nano-technology and the lower temperature range providing more sensitivity, it is not unachievable.

The quantum statistics obeyed by the whole body modes of a nano-oscillator array can also be probed in  an optical scattering experiment, and the results so obtained would be preferable to those from a specific heat measurement, enabling one to observe the behavior of particular modes.   We approximate the scattered field due to thermal vibration of a nano-structured array by estimating the classical electromagnetic scattering from a vibrating structure, and then replacing the required averages over displacement fields in the classical expressions by quantum expectation values.  The scattered component of the vector potential is given by
\begin{equation}
\label{formal_A_field}
A(\vec x, \omega) = {\mu_0 \over 4\pi} \int_V \vec J_{eq}(\vec x') {\exp (ik|\vec x-\vec x'|) \over |\vec x-\vec x'|} d^3 x'
\end{equation}
where the equivalent source current is $\vec J_{eq} = -i\omega (\epsilon -\epsilon_0)\vec E$$ = -i\omega \epsilon_0 \chi \vec E$ with $\epsilon$ the dielectric constant, $\chi$ the susceptability of the system, and  $\vec E$ the total electric field inside the the dielectric.  To leading order, $\vec E$ can be replaced by the incident field in the integration over the source, a Born approximation which is reasonable because of the small thickness of the structure.  

The evaluation of the scattering cross section can be further simplified by the following approximations.  We assume: the motion is slow in comparison with optical frequencies; the displacements and sample thickness are small relative to an optical wavelength; and, finally, the displacements are dominated by the normal components.  One may then show that the cross section is given by
\begin{equation}
\label{semi-cross-section}
{d^2 \sigma \over d\Omega d\omega} = \left ( {k^2 \chi t \over 4\pi}\right )^2  \left [ 2\pi\delta (\Omega ) |a_0 (q)|^2 + q_z^2 |\psi (q, \Omega|^2 \right ] \ .
\end{equation}
Here $q = k - k_i$ is the change of the optical wavenumber and $\Omega$ is the change of the incident frequency.  The quantity $a_0 (q )=\int d^2 x \exp (-iqx)$ is a static aperture integral over the surface and gives the Rayleigh scattering, while $\psi (q,\Omega )$ is the normal displacement field of the system, and gives the dynamic response.  A reasonable approximation to the cross section is then obtained by replacing $|\psi(q, \Omega )|^2$ by the thermal average of the quantum expectation value of Eq. 1,
\begin{eqnarray}
\label{general_expect}
&&<|\psi(q, \Omega )|^2>_T = \int [d\alpha ]   \left [ {\hbar \over 2 \rho \omega_\alpha} \right ] \\
 &&\quad \left \{  \right .2 \left . \pi \delta (\Omega - \omega_\alpha ) n (\omega_\alpha , T) |\psi_\alpha(q, \Omega )|^2 \right . 
 \nonumber \\
&& \quad\quad +  2 \left . \pi \delta (\Omega + \omega_\alpha ) (n (\omega_\alpha , T) + 1) |\psi_\alpha(-q, \Omega )|^2
\right \} \ , \nonumber
\end{eqnarray}
where $n (\omega_\alpha , T)$ is the Bose distribution, and the two terms correspond to Stokes and anti-Stokes peaks respectively in the usual way for Brillouin scattering.  For extended modes, the  wave function $\psi_\alpha$ takes on the simple form, $\psi_{nk} (q) = (2\pi /a)^2 \delta_N^2(q-k) l^2 j_{nk}(q-k)$, where $\delta^2_N (k) = \sum_m \exp(-ik\cdot ma)$ is a lattice delta function, and $j_{nk}(q) = l^{-2}\int d^2 x \exp (-iqx) u_{nk}(x)$ is the modal optical source strength, the normalized Fourier transform of the mode shape of a single element of the array.  

Consider the spectra associated with a phonon of wavenumber $q$ which scatters light at angle $\sin^{-1}(q/2k)$.  Using Eq. (\ref{semi-cross-section}) and Eq. (\ref{general_expect}), we estimate the ratio $R$ of the Brillouin peak to Rayleigh scattered levels for a 100$\mu$m$\times 100\mu$m sample.  For normal incidence, scattering angle $\theta$, and 1GHz phonons, we find $R \sim 4\times 10^{-6}\theta^4$ away from the main Rayleigh lobe in directions along a diagonal ($\theta = \phi = \pi /4$).  Using a multi-pass Fabry-Perot interferometer with contrast ratio $10^9$, the Brillouin peak level is found to equal the background (and thus be readily observable) for $\theta \ge 7\deg$.  

Having established that the Brillouin spectra is observable above the Rayleigh background, we consider the scattered power level.  The total power scattered into solid angle $\Delta \Omega$ near an anti-Stokes peak may be obtained from Eqs. (\ref{semi-cross-section}) and (\ref{general_expect})  as
\begin{equation}
\label{Brillouin_power}
I  \approx I_0  \left (  {r j_{nq}(0) k^2 l^2 \Delta x ( 1+\cos\theta ) \over 2\pi a} \right )^2 
n(\omega_q , T) \Delta \Omega \ .
\end{equation}
Here $I_0$ is the total power incident on the array; $k$ is the optical wavenumber; $r = (kd\chi /2)$ is the plane wave reflection coefficient; $j_{nq}(0)$ is the optical source strength; $\Delta x = (\hbar / 2 \rho l^2 \omega_q )^{1/2}$ is the size of the ground state wave packet of a single element of the array; and  $\Delta \Omega$ is the solid angle in which light is gathered.  The dependence of the scattered power on the extent of the ground state wave function for a single element of the array is a result of coherent scattering associated with the $N$ oscillators, each moving with the average displacement $\Delta x_M = (\hbar / 2 N \rho l^2 \omega_q )^{1/2}$.  The peak scattered Brillouin power at low temperatures therefore scales as $N$.

Note also that the scattered power is proportional to the mean number $n(\omega_q , T)$ (or  $(n(\omega_q , T) + 1)$ for the Stokes peak) of phonons present at temperature $T$.  Therefore, a measurement of the temperature dependence of the Stokes (anti-Stokes) peak would provide direct evidence that the mode $q$ is obeying quantum statistics.  This requires measurements in the temperature range $kT \sim \hbar \omega_q$.

We now give an order of magnitude estimate of the power scattered into a Stokes peak for the geometry previously considered at low temperatures.
Using the same parameters as employed above and, further, assuming $j_{nk} =  \Delta \Omega = 1$, we find $I \sim 4\times 10^{-14}I_0$ at low temperatures such that $n(\omega_q , T) = 1$.  Taking $I_0 = 100\mu{\rm W}$, a heat load well within the cooling capacity of many current systems, we obtain a Brillouin scattered power of approximately 10 photons/sec.  This is indeed a small signal, but note that the power level $I$ in Eq. (\ref{Brillouin_power}) scales as $\lambda^2$ if we rescale the nano-structure, $L \rightarrow \lambda L$, provided we do not violate the small dimension approximations.  In obtaining this scaling law, we have assumed the incident power to be fixed; if, instead, the power flux is regarded as fixed, an additional factor of $\lambda^2$ results.  Thus, higher signal levels will be observed for larger structures, at the cost of lower temperatures.

Topological disorder in the structure reduces the coherence of the Bloch modes and thus the levels of the Stoke's and anti-Stoke's peaks.  Near band edges, however, the Bloch modes of the array are localized by irregularity in the system\cite{sheng-john_localization}, and the higher resulting vibration levels partially compensate for the loss of coherence.  The mean square displacement level associated with quantum level motion increases as $\xi^{-2}$ as the localization length $\xi$ decreases.  Assuming the $q = 2|k| \sin\theta /2$ selection rule breaks down as a result of the localization,  as in the case of amorphous materials\cite{Shuker}, all $q$'s contribute to the scattering at each angle, and the strength of the scattering is proportional to the area of mode coherence $l^2$($l$ is the coherence length).    The cross section for Brillouin scattering will, therefore, be proportional to $(l/\xi )^2 n(\omega )$, where $n(\omega )$ is the density of states.  For strong localization of a resonance near a band edge, $l/\xi$ is of order 1, and $n(\omega )$ is of order $N$, leading to a scattered power of the same order of magnitude as given in Eq. (\ref{Brillouin_power}).  This phenomena should be observable near the band edges associated with the gaps at 1.5GHz and 6GHz for the geometry considered here.

We have explored the extent to which the quantum behavior of mechanical oscillators can be probed by observing  two of the simplest physical properties of a large, free standing array of coupled nano-oscillators: specific heat and Brillouin scattering.  In both cases, the measurements are challenging but within reach of current capability.  Carrying out such experiments at low temperatures would enable us to observe the quantum statistics obeyed by rigid body modes involving large numbers of atoms and, thus, verify that the energy levels of these states are quantized.  Such measurements would significantly enlarge the domain in which quantum mechanical predictions have been tested experimentally.  We have not examined the possibility of employing such a system to explore coherent quantum phenomena, but because the observation of small occupation number states is not out of reach, such an attempt seems promising.

The authors gratefully acknowledge the Office of Naval Research for support of this work.

\bibliography{/Users/dougp/Documents/acoustics}

\end{document}